# Detecting Malignant TLS Servers Using Machine Learning Techniques

Sankalp Bagaria, R. Balaji, B. S. Bindhumadhava
Centre for Development of Computing, Bangalore, India

Email: {sankalp, balaji, bindhu}[at]cdac[dot]in

*Abstract*-TLS uses X.509 certificates for server authentication. A X.509 certificate is a complex document and various innocent errors may occur while creating/ using it. Also, many certificates belong to malicious websites and should be rejected by the client and those web servers should not be visited. Usually, when a client finds a certificate that is doubtful using the traditional tests, it asks for human intervention. But, looking at certificates, most people can't differentiate between malicious and non-malicious websites. Thus, once traditional certificate validation has failed, instead of asking for human intervention, we use machine learning techniques to enable a web browser to decide whether the server to which the certificate belongs to is malignant or not ie, whether the website should be visited or not.

Once a certificate has been accepted in the above phase, we observe that the website may still turn out to be malicious. So, in the second phase, we download a part of the website in a sandbox without decrypting it and observe the TLS encrypted traffic (encrypted malicious data captured in a sandbox cannot harm the system). As the traffic is encrypted after Handshake is completed, traditional pattern-matching techniques cannot be employed. Thus we use flow features of the traffic along with the features used in the above first phase. We couple these features with the unencrypted TLS header information obtained during TLS Handshake and use these in a machine learning classifier to identify whether the traffic is malicious or not.

## I. INTRODUCTION

Public Key Infrastructure solves the problem of distributing keys to the servers. It uses the hierarchy of certifying authorities to distribute and manage the keys using the certificate-chain. A X.509 certificate is used by the server to prove to the client that a particular public key belongs to the server. The client wanting to connect to a server securely requests its TLS certificate. When server sends its certificate, the client verifies it and proceeds to exchange information with the server through encrypted records.

Transport Layer Security (TLS) is the most popular protocol to encrypt the network traffic. TLS is the greatest user of X.509 certificates, especially for server authentication. TLS has two phases: Handshake Phase and Data Transfer Phase. In Handshake phase, various algorithms and parameters required for secure data transfer are negotiated. Once handshake is complete and symmetric keys are agreed upon, application data is transferred between client and server in the form of encrypted records.

Studies show that as much as 60 % of network traffic uses TLS. When a browser finds a certificate that is suspicious, it raises a warning to the human user. Then, human may accept or reject the certificate. But, most users are not able to differentiate between benign certificates and malicious certificates. So, we try to identify why traditional certificate validation has failed and whether the site is really malicious or it is innocent. We use machine learning techniques to decide this and whether the website should be visited or not.

Even when the site passes the above test, the site may still turn out to be malicious. So, in the second phase, a part of the website data is downloaded in a sandbox and analysed through machine learning techniques without decrypting it. Encrypted malicious traffic can't harm the user's system. It is discarded if found malicious. And, it is processed and displayed if found not to be malicious.

The remainder of the paper is organized as follows: Section II deals with traditional certificate validation. It talks about how both innocent mistakes and malicious errors may occur while creating /using X.509 certificates. Section III reviews the feature set used for identifying malicious certificates in the first phase. Section IV reviews the feature set used for identifying malicious traffic in the second phase. Together, these two phases decide if the server is malignant or not. Section V reviews previous and related work. Section VI discusses limitations and future work. Section VII concludes the paper and Section VIII cites the acknowledgments.

## II. TRADITIONAL CERTIFICATE VALIDATION: INNOCENT MISTAKES AND MALICIOUS ERRORS

X.509 Certificate is complex document. So, innocent mistakes crop while creating/ using the document. Some of the innocent mistakes that may occur in the certificate are also found to occur in malicious certificates resulting in various attacks. Traditional certificate validation cannot differentiate between the two. So, they leave it to the user to decide whether to visit the website or not. Innocent mistakes and malicious errors because of which traditional certificate validation may fail are described as follows:

A. **KeyUsage**: KeyUsage field of X. 509 certificate defines the scope of the certificate as to for what all purposes it may be used.

```
KeyUsage ::= BIT STRING {
        digitalSignature    (0),
        nonRepudiation      (1),
        keyEncipherment     (2),
        dataEncipherment    (3),
        keyAgreement        (4),
        keyCertSign         (5),
        cRLSign             (6),
        encipherOnly        (7),
        decipherOnly        (8) }
```

CAs make mistakes while making certificates. Moreover, many software don't cross-check KeyUsage field while validating certificates. KeyUsage field has been used in attacks like described in [6] and [7].

B. **Validity Dates**: The certificate is valid within a specified period given by notBefore and notAfter fields.

```
Validity ::= SEQUENCE {
      notBefore    Time,
      notAfter     Time }
```

Sometimes, an administrator may forget to renew the certificate. So, if the certificate has expired recently, it's possible that the certificate is not malicious.

C. **Critical Extensions**: To preserve backward compatibility, it is considered ok if the client doesn't understand the extension but should reject the certificate, if it is marked critical. Both innocent mistakes and malicious errors are common while filling critical extensions.

```
Extension ::= SEQUENCE {
        extnID        OBJECT IDENTIFIER,
        critical      BOOLEAN DEFAULT FALSE,
        extnValue     OCTET STRING
                        - contains the DER encoding of an ASN.1 value
                        - corresponding to the extension type identifed
                        - by extnID
}
```

D. **Hostname Validation**: Although an innocent user can make a mistake, while creating the certificate, in hostname validation eg, not checking subject and subject alternative fields properly (eg, paypai.com instead of paypal.com) or using more than one level of wildcard or not putting it at leftmost position; an attacker can also launch impersonation attack using these certificates eg, often, browsers will just verify if the `subject Name` contains the string say, `mywebsite.com`, or will use a regex that either accepts `mywebsite.com.evil.com` or such evil subdomains.

```
TBSCertificate ::= SEQUENCE {
        version       [0] EXPLICIT Version DEFAULT v1,
        serialNumber  CertificateSerialNumber,
        signature     AlgorithmIdentifier,
        issuer        Name,
        validity      Validity,
        subject       Name,
```

E. **Basic Constraints**: The BasicConstraints extension specifies if a certificate is that of a CA and how many intermediate CAs can follow it before a leaf certificate. Improper/ Incomplete filling of the Basic Constraints in the certificate is a common mistake committed by a CA.

Id-ce-basicConstraints OBJECT IDENTIFIER ::= { id-ce 19 }

```
BasicConstraints ::= SEQUENCE {
        cA                BOOLEAN DEFAULT FALSE,
        pathLenConstraint INTEGER (0..MAX) OPTIONAL }
```

[8] describes some of the attacks that were launched. One of the attacks involved a tool called sslsniff that given a valid leaf-node certificate for any domain, could generate certificate for the domain, client was connecting to and sign it using the leaf-node. Thus, only CAs must be allowed to sign the certificates and pathLen has to be verified.

F. **Name Constraints**: The NameConstraints extension sets limitations for the certificates that can follow in the CA's chain. Not filling it properly is a common mistake made by users while creating certificates.

Id-ce-nameConstraints OBJECTIDENTIFIER ::= { id-ce 30 }

```
NameConstraints ::= SEQUENCE {
        permittedSubtrees  [0]   GeneralSubtrees OPTIONAL,
        excludedSubtrees   [1]   GeneralSubtrees  OPTIONAL }
```

GeneralSubtrees ::= SEQUENCE SIZE (1..MAX) OF GeneralSubtree

```
GeneralSubtree ::= SEQUENCE {
        base            GeneralName,
        minimum   [0]   BaseDistance DEFAULT 0,
        maximum   [1]   BaseDistance OPTIONAL }

BaseDistance ::= INTEGER (0..MAX)
```

According to [9], a subordinate CA of ANSSI issued an intermediate certificate that they installed on a network monitoring device, which enabled the device to act as a MITM of domains or websites that the certificate holder did not own or control. In early 2011, a hacker hacked the DigiNotar CA and issued certificates for *.google.com, *.skype.com and *.*.com, as well as few intermediate CA certificates carrying the names of well-known roots. The *.google.com certificate was used to launch a MITM attack against Gmail users in Iran. That is, the attackers were able to create both CA and leaf certificates through an existing CA. [16] describes this attack.

### III. FEATURE SELECTION FOR IDENTIFYING MALICIOUS CERTIFICATE IN FIRST PHASE

A Decision Tree (C4.5 classifier is used to separate malicious certificate from benign X.509 certificates. 66% data is used to train the classifier and the rest of the data is used for testing. Following features are chosen after doing numeric-to-nominal conversion. Depending on the values of the features, the classifier gives a boolean output that the certificate is malicious or benign. This is used by the browser to decide whether to visit the website or not. Features chosen are as follows:

**1. Reason(s) for the Certificate Failing the Traditional Certificate Validation:** This step has occured because the certificate has failed one or many criteria used in traditional certificate validation. As explained above, these criteria are KeyUsage, Validity Dates, Critical Extensiosn, Hostname Validation, Basic Constraints, Name Constraints. Boolean values are assigned to each of these criteria. If certificate validation fails because of a particular criterion, the value for that criterion is True (1) otherwise it is False (0). As discussed, a particular criterion may fail for malicious or innocent reasons. Please note that certificate may fail for one or more than one reasons.

2. **Self-Signed Certificates**: Self-signed certificates can be employed by legitimate servers but it remains a fact that many malicious sites use self-signed certificates. A study [10] shows that on an average, only 0.09% legitimate TLS sessions use self-signed certificate while 0.7% malicious sessions use self-signed certificates, an order of magnitude higher than legitimate sessions. Thus, if a certificate is self-signed, chances of it being malignant is higher by an order of magnitude. Thus, it is chosen as a feature for our Net-Bayesian classifier.

### IV. FEATURE SELECTION FOR IDENTIFYING MALICIOUS TRAFFIC IN SECOND PHASE

Net-Bayesian classifier is used to separate malicious traffic from a malignant TLS server from legitimate TLS traffic from a genuine server. Net Bayesian classifier is used because features are not independent. 66% data is used to train the classifier and the rest of the data is used for testing. Following features are chosen after doing numeric-to-nominal conversion Depending on the values of the features, the classifier gives a boolean output that the TLS encrypted traffic is malicious or legitimate. This is used by the browser to decide whether to download the full web page(s) or not. Features chosen are as follows:

1. **Features of Classifier of Phase 1:** The above features used in Phase 1 are also used in Phase 2. They are the reasons for the certificate failing the traditional certificate validation and whether the server certificate is self-signed.

2. **Flow Metadata:** Traditional flow data are the first set of additional features for the classifier. They are the number of inbound bytes, outbound bytes, inbound packets, outbound packets; the source and destination ports; and the total duration of the flow in seconds.

3. **Packet Lengths and Packet Inter – Arrival Times:** Minimum, Maximum, Mean and Standard Deviation of Packet Lengths and Minimum, Maximum, Mean and Standard Deviation of Packet Inter – Arrival Times during the duration of flow are taken as the second set of additional features for the classifier.

4. **Unencrypted TLS Header Information exchanged during TLS Handshake:**
4a. **Critical extensions:** Malicious servers rarely select TLS extensions. Legitimate servers select different TLS extensions. 0Xff01 (renegotiation_info) and 0x000b (ec_point_formats) are most common. Usually, 21 unique extensions are observed, most of them in legitimate traffic. A binary vector of length 21 was created with a true (1) if extension is present and a false (0) if it is absent.

4b. **Weak ciphersuite:** Approx 90% of the malicious servers use one of the following ciphersuite: 0x000a (TLS_RSA_WITH_3DES_EDE_CBC_SHA), 0x0004 (TLS_RSA_WITH_RC4_128_MD5), 0x006b (TLS_DHE_RSA_WITH_AES_256_CBC_SHA256) and 0x0005 (TLS_RSA_WITH_RC4_128_SHA). TLS_RSA_WITH_RC4_MD5 and TLS_RSA_WITH_RC4_128_SHA are considered weak. A numeric value is assigned to ciphersuite to identify which ciphersuite server will be used. It helps identify malicious traffic.

## V. RELATED WORK

[1] RFC 5246 describes Transport Layer Security (TLS) Protocol Version 1.2 Version 1.3 is being developed. [2] gives the latest internet-draft of TLS 1.3. [3] RFC 7457 summarizes known attacks on TLS and DTLS. Details of X.509 Certificate can be found in RFC 5280. It specifies how a certificate should look like and be used.

[5] discusses various pitfalls committed while creating certificates. [10] discusses how malware uses TLS and shows how data features from passive monitoring of TLS can be used for malware identification and family attribution. [14], [15], [16] and [17] specify various details of TLS errors occurring in the World Wide Web.

The errors generated by traditional certificate validation are used by our classifier along with other features to classify a certificate as malicious or non-malicious. This is our unique idea. The browser needs to be enabled with this. If the certificate manages to deceive the first phase, the encrypted traffic with Header Information along with the phase 1's features are analysed and a decision is made if the website is malicious or not.

Thus the decision whether a site is malicious or not and should be visited or not is shifted to the browser. We have done this because most people are not able to differentiate between malicious and innocent websites without visiting it, by using the data available to them when a certificate fails the traditional certificate validation. We can train the classifier to be extra-careful during training by giving more legitimate traffic.

## VI. LIMITATIONS AND FUTURE WORK

We used C4.5 Classifier for the first phase and Net-Bayesian classifier for the second phase. Various other machine learning classifiers like boosted classifiers have to be implemented and tested so that the best possible results can be obtained. For phase 1, our feature set consisted of reasons for the certificate failing the traditional certificate validation and a boolean variable citing if the certificate is self-signed or not. For phase 2, our feature set consisted of feature set of phase 1, flow metadata, packet length and packet inter-arrival time and; TLS Header information exchanged during TLS Handshake was also included. It has to be observed if more features eg, byte distribution will improve the performance for the classifier. Finally, this idea has to be included in the browsers.

## VII. CONCLUSIONS

60% traffic on Internet is on TLS. TLS uses X.509 certificates for server authentication. While designing a complex document as X.509 certificate, both genuine innocent errors and malicious errors may occur. Usually, when a browser finds a certificate that is doubtful using the traditional certificate validation, it asks for human intervention. But most people can't differentiate between malicious and non-malicious certificates. Thus, in this paper, we proposed that once traditional certificate validation has failed, instead of asking for human intervention, machine learning techniques be employed to enable a web browser to decide whether the server to which the certificate belongs to, is malicious or not ie, whether the website should be visited or not.

We also observed that once a certificate has been accepted in the above phase, the web server may still turn out to be containing malicious content. So, in the second phase, we download part of website in a sandbox without decrypting it and observe the TLS encrypted traffic (encrypted malicious data captured in a sandbox cannot harm the system). As the traffic is encrypted after Handshake is completed, traditional pattern-matching techniques cannot be employed. Thus we used flow features of the traffic along with the features used in the above first phase. We couple these features with the unencrypted TLS header information obtained during TLS Handshake and use these in a machine learning classifier to identify whether the traffic from the TLS server is malicious or not.

## VIII. ACKNOWLEDGEMENTS